# Using Non-Lipschitz Signum-based Functions for Distributed Optimization and Machine Learning: Trade-off Between Convergence Rate and Optimality Gap

**Mohammadreza Doostmohammadian [1,*], Amir Ahmad Ghods [1], Alireza Aghasi[2], Zulfiya R. Gabidullina[3] and Hamid R. Rabiee[4]**

[1] Faculty of Mechanical Engineering, Semnan University, Semnan, Iran; doost@semnan.ac.ir (M Doostmohammadian); amir-ghods@semnan.ac.ir (AA Ghods)
[2] Department of Electrical Engineering and Computer Science, Oregon State University, Oregon, USA; alireza.aghasi@oregonstate.edu
[3] Institute of Computational Mathematics and Information Technologies, Kazan Federal University, Russia; Zulfiya.Gabidullina@kpfu.ru
[4] Computer Engineering Department, Sharif University of Technology, Tehran, Iran; rabiee@sharif.edu
* Correspondence: doost@semnan.ac.ir

**Abstract:** In recent years, the prevalence of large-scale data-sets and the demand for sophisticated learning models have necessitated the development of efficient distributed machine learning (ML) solutions. Convergence speed is a critical factor influencing the practicality and effectiveness of these distributed frameworks. Recently, non-Lipschitz continuous optimization algorithms have been proposed to improve the slow convergence rate of the existing linear solutions. The use of signum-based functions is previously considered in consensus and control literature to reach fast convergence in the prescribed time and also to provide robust algorithms to noisy/outlier data. However, as shown in this work, these algorithms lead to an optimality gap and steady-state residual of the objective function in discrete-time setup. This motivates us to investigate the distributed optimization and ML algorithms in terms of trade-off between convergence rate and optimality gap. In this direction, we specifically consider the distributed regression problem and check its convergence rate by applying both linear and non-Lipschitz signum-based functions. We check our distributed regression approach by extensive simulations. Our results show that although adopting signum-based functions may give faster convergence, it results in large optimality gaps. The findings presented in this paper may contribute to and advance the ongoing discourse of similar distributed algorithms, e.g., for distributed constrained optimization and distributed estimation.

## 1. Introduction

Distributed algorithms for detection, estimation machine learning and resource allocation have recently gained interest in signal processing, control, and optimization literature [1-4]. Such algorithms are known to have many benefits in terms of scalability, real-time and parallel data processing, and distributed learning over multi-agent networks, with specific applications in data-mining [5,6]. To give more details, by distributing the computational load across multiple nodes or devices, these algorithms allow for the processing of large data-sets and complex tasks without overburdening a single central processing unit [7]. This is also motivated by recent Cloud-based solutions for high-performance computing, which are well-suited for computationally intensive tasks involved in distributed algorithms. This is especially beneficial in machine learning and simulation-based algorithms where large data-sets and complex computations are common. Such distributed algorithms are further motivated by recent advances in Internet-of-Things (IoT) applications [8], Cloud/Edge computing [9], and Cyber-Physical Systems (CPS) [10,11] with many processing devices interconnected over a network. In general, distributed optimization algorithms benefit from *robustness* to single-node-failure, *parallelism* by assigning different parts of the optimization process to different computing nodes/agents, *scalability* (and resource utilization) by distributing the workload for large-scale optimization problems, and *data distribution* without the need for centralized storage. In this work, we investigate the use of non-Lipschitz consensus-based functions for distributed optimization in terms of convergence rate and steady-state optimization residual (optimality gap).

### *1.1. Literature Review:*

Signum-based functions can be robust to outliers in the data. In the presence of noisy or outlier data, algorithms based on the signum function may be more resilient compared to methods that rely on smooth functions [12,13]. In the consensus literature these algorithms are used to reach fast agreement over the multi-agent network in prescribed-time [14], finite-time [15,16], and fixed-time [17,18]. For similar reason, signum-based functions are used for distributed resource allocation [19], event-triggered systems [20], distributed gradient-flow schemes [21], sliding-mode control [22], parameter identification [23], distributed optimization [24], cooperative control [25,26], and distributed estimation [13]. A survey of finite/fixed-time convergent algorithms can be found in [27].

In the context of distributed optimization, ML, and regression solutions, most existing literature provide linear algorithms [28-30], many of which are consensus-based solutions [31,32]. Recently distributed non-Lipschitz algorithms are proposed in the context of distributed optimization and learning, claiming improved convergence rate (stability in finite-time, fixed-time, or prescribed-time) [33-41]. Although in continuous-time these may work properly, the discrete-time dynamics (after discretization) results in steady-state oscillation around the optimal point, known as *chattering* phenomena. This is well-known in nonlinear control applications, e.g., in sliding mode control [42]. For distributed optimization, this results in the final residual of the objective function and optimality gap. In this direction, the current study investigates the trade-off between the improved convergence rate and the optimization steady-state residual.

The use of non-Lipschitz function also includes non-Lipschitz activation functions in neural networks [43,44] and non-Lipschitz optimization in related fields such as deep learning [45,46] by designing algorithms that exploit structure, e.g., sign/threshold dynamics and proximal maps for nonconvex nonsmooth terms. In neural networks, while mainly Lipschitz activations (ReLU, ELU, GELU) is adopted, classical examples include the signum function and hard-threshold activations.

### *1.2. Contributions:*

This study investigates the distributed optimization algorithms with signum-based functions added to improve their convergence rate. We consider a gradient-tracking (GT) distributed optimization algorithm which is based on consensus algorithm [47]. The idea is to improve the rate of convergence by adding non-Lipschitz sign-based functions, while checking the steady-state residual (the optimality gap). The adopted signum-based functions are sign-preserving and odd, therefore, not violating the consensus-type nature of the algorithm. Further, the GT-based dynamics ensures evolution of the states toward the optimal point. We specifically show that the optimal point is the invariant state under the proposed dynamics. As an ML application, we consider distributed linear regression over a randomly-generated dataset. Our results show that, although sign-based nonlinearity may affect the convergence rate (and reach fixed-time, finite-time, or prescribed-time convergence), it may cause steady-state residual in the cost and certain optimality gap depending on the parameters of the signum-based function.

*1.3. Paper organization:*

Section 2 states the preliminaries. Section 3 frames the distributed linear regression as a distributed optimization problem. Section 4 presents the linear and signum-based GT solutions. Section 5 provides the simulation, and Section 6 concludes the paper.

## 2. Preliminaries

*2.1. Notations*

Let $\lambda$ denote the eigenvalue. The abbreviations LHP and RHP correspond to left-half-plane and right-half-plane in the complex eigenspace. Let $\partial_t z = \frac{dz}{dt}$ be the derivative with respect to $t$. $\mathbf{1}_n^\top \coloneqq \underbrace{(1,\cdots,1)}_{n}$ $\mathbf{0}_n^\top \coloneqq \underbrace{(0,\cdots,0)}_{n}$, i.e., $\mathbf{1}_n$ and $\mathbf{0}_n$ are size $n$ vectors of all 1's and 0's. The operator $';'$ implies column concatenation of vectors. $\nabla F$ is the gradient of $F$.

*2.2. Algebraic Graph Theory*

The distributed algorithm works over a connected undirected multi-agent network represented by an undirected graph topology $\mathcal{G}$ with real adjacency matrix $W = [w_{ij}] \in \mathbb{R}^{n\times n}$. The entry $w_{ij} > 0$ associated with the link $j \to i$ denotes the weighting factor, which defines the weight that agent $i$ assigns to the information received from agent $j$. For a connected undirected $\mathcal{G}$, its associated matrix $W$ is irreducible. Further, define the Laplacian matrix $\overline{W} = [\overline{w}_{ij}] \in \mathbb{R}^{n\times n}$ as $\overline{w}_{ij} = w_{ij}$ for $i \neq j$ and $\overline{w}_{ij} = -\sum_{i=1}^{n} w_{ij}$ for $i = j$. It is known that the connectivity of the graph is related to the rank of its Laplacian matrix. Given a connected undirected graph $\mathcal{G}$, its Laplacian $\overline{W}$ has only one zero eigenvalue and the rest are on LHP. The (left and right) eigenvectors $\mathbf{1}_n^\top$ and $\mathbf{1}_n$ are associated with this zero eigenvalue, i.e., $\mathbf{1}_n^\top \overline{W} = \mathbf{0}_n$ and $\overline{W}\mathbf{1}_n = \mathbf{0}_n$ [47].

*2.1. Background on Signum-based Consensus*

Consensus algorithms are widely used to coordinate (reach agreement) over multi-agent networks. The primary solution to solve consensus in a distributed way is to follow a linear dynamics, where the dynamics at node $i$ is as follows [47]:

$$\dot{x}_i = -\eta_1 \sum_{j\in N_i} w_{ij}(x_i - x_j), \tag{1}$$

where $\eta_1 > 0$ as the step-rate and $W = [w_{ij}]$ as the stochastic adjacency weight matrix[1], $N_i$ denotes the neighboring set of node/agent $i$, and $x_i, x_j$ denote the state values at nodes $i, j$. It is known that, under certain conditions, the solution of this dynamics asymptotically converges to the agreement/consensus state. On the other hand, finite-time consensus protocols [16,48] improve the convergence rate of the linear dynamics (1) in the region $|x_i - x_j| < 1$ by adding signum-based function as follows:

[1] A matrix is called stochastic if for every $i \in \{1,\dots,n\}$ we have $\sum_{j=1}^{n} w_{ij} = \sum_{j=1}^{n} w_{ji} = 1$.

$$\dot{x}_i = -\eta_1 \sum_{j \in N_i} w_{ij} sgn^{v_1}(x_i - x_j), \quad (2)$$

where $0 < v_1 < 1$ and the signum-based function $sgn^{v_1}(x): \mathbb{R} \to \mathbb{R}$ is

$$sgn^{v_1}(x) = x\,|x|^{v_1 - 1}, \quad (3)$$

recall that this function is non-Lipschitz at $x = 0$. As it is proved in finite-time consensus literature [16,48], this solution converges faster than linear dynamics (1) in the regions close to the equilibrium. This follows from the definition of signum-based function and the fact that $|sgn^{v}(x)| > |x|$ for all $|x| < 1$. Moreover, the non-Lipschitz continuity and infinite gradient at the agreement equilibrium allow to reach consensus in finite time. However, these solutions converge slower than linear dynamics (1) in the region farther from zero (or the agreement equilibrium) as $|sgn^{v}(x)| < |x|$ for all $|x| > 1$. Fixed-time consensus protocols [49-53] overcome this by adding a second term in the form $sgn^{v_2}(x)$ with $v_2 > 1$. For this function we have $|sgn^{v_2}(x)| > |x|$ for $|x| > 1$; this implies a faster convergence rate than the linear dynamics for states farther from the agreement equilibrium. By combining the two consensus dynamics, the solution has a fast convergence rate for all the regions, both close and far from the equilibrium. This convergence rate can be changed via the parameters $v_1$, $v_2$. The overall fixed-time consensus dynamics is in the form:

$$\dot{x}_i = -\sum_{j \in N_i} w_{ij} \left( \eta_1 sgn^{v_1}(x_i - x_j) + \eta_2 sgn^{v_2}(x_i - x_j) \right), \quad (4)$$

with $0 < v_1 < 1$, $v_2 > 1$, $\eta_2$, $\eta_1 > 0$. The convergence rate of the dynamics (4) is faster than protocols (1) and (2). It should be mentioned that these dynamics are *non-Lipschitz*[2]. Therefore, they may result in chattering around the equilibrium as stated in sliding-mode control literature [42].

## 3. The Framework: Distributed Regression Problem

Linear regression is a statistical method used to fit a linear line to a set of data points. Given a set of $N$ data points $\chi_i \in \mathbb{R}^{m-1}, i = \{1, \dots, n\}$, the model's prediction is $\beta^T \chi_i - \nu = y_i$, which gives the hyperplane that fits the data best. In the centralized regression, all the data points are sent to a central computation entity (the fusion center) which finds $[\nu;\ \beta]$ optimizing the following quadratic convex function:

$$\min_{[\nu;\beta]} \sum_{i=1}^{N} (\beta^T \chi_i - \nu - y_i)^2, \quad (5)$$

which is also known as the linear least square problem. Distributed Linear Regression (DLR), is an extension of linear regression that leverages distributed computing resources for handling large datasets. In traditional linear regression, all data is typically processed on a single machine (the fusion center), which can become impractical when dealing with massive datasets that may not fit into the memory of a single computer. Distributed linear regression distributes the computation across multiple machines or nodes in a computing cluster. For parallel processing of data, this approach allows one to make handling large-scale datasets more feasible (and more efficient). In DLR, the dataset is widespread over a network of $n$ agents/machines, and each machine $i$ has its own $\frac{N}{n} \leq N_i \leq N$ data points $\chi^i$, where some of these data might be shared between two or more machines. The main idea is to solve the optimization problem (5) locally at each machine using its own data $\chi^i$ and information received from its neighbor machines. Note that every machine has access to partial data and thus the optimal values $\beta_i$ and $\nu_i$ may differ for each machine $i$. Therefore, the machines share necessary information by reaching a consensus on $\beta$ and $\nu$. then, the optimization problem changes to:

$$\min_{\beta_1,\nu_1,\dots,\beta_n,\nu_n} \sum_{i=1}^{n} f_i(\beta_i,\nu_i), \quad\quad (6)$$
$$subject\ to\ \beta_1 = \cdots = \beta_n,\ \nu_1 = \cdots = \nu_n$$

$$f_i(\beta_i,\nu_i) = \sum_{j=1}^{N_i} \left(\beta_i^T \chi_j^i - \nu_i - y_j\right)^2, \quad\quad (7)$$

[2] A function $f(x): x \in \mathbb{R}$ is called Lipschitz continuous if there exists a constant $\mathcal{K}$ such that for every two points $x_1, x_2$ we have, $|f(x_1) - f(x_2)| \leq \mathcal{K}|x_1 - x_2|$. Otherwise, the function is called non-Lipschitz.

This problem (6)-(7) represents a consensus-constrained distributed optimization framework. By denoting the optimization state variable as vector $x_i = [\beta_i^T; \nu_i]$ and vector $x$ be the concatenation of all local state vectors $x_i$'s, i.e., $x = [x_1; x_2; \dots; x_n] \in \mathbb{R}^{mn}$. Then, this DLR problem (6)-(7) is framed as a distributed optimization formulation:

$$\min_{x} \quad F(x) = \sum_{i=1}^{n} f_i(x_i), \quad\quad (8)$$
$$subject\ to\ x_1 = x_2 = \cdots = x_n$$

## 4. The Proposed Signum-based Learning Dynamics

### *4.1. The Algorithm*

First, we recall the existing linear gradient-tracking dynamics to solve the DLR asymptotically. The following linear dynamics is proposed by the author in [7] to solve distributed optimization and support-vector-machine problems:

$$\dot{x}_i = -\sum_{j=1}^{n} w_{ij}\left(x_i - x_j\right) - \alpha y_i, \quad\quad (9)$$

$$\dot{y}_i = -\sum_{j=1}^{n} a_{ij}\left(y_i - y_j\right) - \partial_t \nabla f_i\,(x_i), \quad\quad (10)$$

with $x_i(t)$ and $y_i(t)$ respectively denoting the state and the auxiliary variable at agent $i$ at time $t$. The auxiliary variable $y_i(t)$ tracks and accumulates the sum of the local gradients. The matrices $A = [a_{ij}] \in \mathbb{R}^{n\times n}$ and $W = [w_{ij}] \in \mathbb{R}^{n\times n}$ are the adjacency weight matrices associated with the state $x$ and auxiliary variable $y$. For convex objective functions, the convergence rate of the dynamics (9)-(10) toward the optimal point is $O\left(exp(\lambda_2 t)\right)$ with $\lambda_2$ as the smallest nonzero eigenvalue of the Laplacian matrix associated with matrices $A$ and $W$ (also known as the algebraic connectivity) [54]. Recalling the application of signum-based functions from Section 2.3, the convergence rate can be improved by using signum-based functions, and the accelerated version of the linear dynamics (9)-(10) is in the following form:

$$\dot{x}_i = -\sum_{j=1}^{n} w_{ij}\left(sgn^{\nu_1}\left(x_i - x_j\right) + sgn^{\nu_2}\left(x_i - x_j\right)\right) - \alpha y_i, \quad\quad (11)$$

$$\dot{y}_i = -\sum_{j=1}^{n} a_{ij} \left( sgn^{v_1}(y_i - y_j) + sgn^{v_2}(y_i - y_j) \right) + sgn^{u_1}(\partial_t \nabla f_i(x_i)) + sgn^{u_2}(\partial_t \nabla f_i(x_i)), \tag{12}$$

with[3] $0 < u_1 < 1,\ u_2 > 1,\ 0 < v_1 < 1,\ v_2 > 1$. The fast convergence of ML dynamics (11)-(12) includes one step of consensus on the states and one step of gradient tracking update. Note that the nonlinear signum-function $sgn^{v_1}(\cdot)$ is odd, sign-preserving, and monotonically increasing; therefore, the stability properties hold similar to the linear case [7]. These properties of $sgn^{v_1}(\cdot)$ function also ensure that:

$$\sum_{i=1}^{n} \dot{y}_i = \sum_{i=1}^{n} sgn^{u_1}(\partial_t \nabla f_i(x_i)) + sgn^{u_2}(\partial_t \nabla f_i(x_i)), \tag{13}$$

$$\sum_{i=1}^{n} \dot{x}_i = -\alpha \sum_{i=1}^{n} y_i\,, \tag{14}$$

these follow the stochastic property of the matrices $W$ and $A$ and the existing consensus-based distributed algorithms (see [47,55] as an example) saying that,

$$\sum_{i=1}^{n}\sum_{j=1}^{n} w_{ij} \left( sgn^{v_1}(x_i - x_j) + sgn^{v_2}(x_i - x_j) \right) = 0, \tag{15}$$

$$\sum_{i=1}^{n}\sum_{j=1}^{n} a_{ij} \left( sgn^{v_1}(y_i - y_j) + sgn^{v_2}(y_i - y_j) \right) = 0\,,$$

by initializing as $y(0) = 0$ it is straightforward to see that $\sum_{i=1}^{n} y_i$ tracks a nonlinear signum-based function of $-\sum_{i=1}^{n} \nabla f_i(x_i)$. This implies that the time-derivative of $\sum_{i=1}^{n} x_i$ tracks a function of the accumulated gradient at all nodes over the network. This follows from the proposed nonlinear structure of (11)-(12) and can be extended to any form of odd sign-preserving model nonlinearities (e.g., quantization or saturation) while preserving the GT dynamics. This is in contrast to the existing linear alternating direction method of multipliers (ADMM) dynamics that does not allow one to consider model nonlinearity. In other words, the existing ADMM solutions [56-59] cannot directly address typical real-world model nonlinearity (such as saturation and quantization), while our proposed dynamics (11)-(12) can address such models. Our distributed solution is summarized in Algorithm 1. For the algorithm initialization, set the $x$ states randomly such that $x(0) \notin \text{span}\{\mathbf{1}_n \otimes \varphi\}$ (with $\varphi \in \mathbb{R}^m$ and $\otimes$ as Kronecker network product) and set $y(0) = \mathbf{0}_{nm}$. From the strict convexity of the DLR objective function $F(x)$ one can see that at the optimal point $x = x^* = \mathbf{1}_n \otimes \overline{x}^*$ (i.e., $x_i = \overline{x}^*$) the equilibrium uniquely holds as,

$$\sum_{i=1}^{n} \dot{x}_i = -\alpha(\mathbf{1}_n^\top \otimes I_m) \left( sgn^{u_1}(\nabla F(x^*)) + sgn^{u_2}(\nabla F(x^*)) \right) = \mathbf{0}_m, \tag{16}$$

which follows from the oddness of the signum function. Similarly, from the proposed dynamics one can see that $x_i = x_j = \overline{x}^*$ and the gradient tracking term $y_i = \mathbf{0}_m$ at the optimal point; thus, we have $\dot{x}_i = \mathbf{0}_m$ and,

$$\begin{aligned} \dot{y}_i &= sgn^{u_1}\left(\partial_t \nabla f_i(\overline{x}^*)\right) + sgn^{u_2}\left(\partial_t \nabla f_i(\overline{x}^*)\right) \\ &= sgn^{u_1}(\nabla^2 f_i(\overline{x}^*)\dot{x}_i) + sgn^{u_2}(\nabla^2 f_i(\overline{x}^*)\dot{x}_i) = \mathbf{0}_m, \end{aligned} \tag{17}$$

[3] Note that for $u_1 = 1,\ u_2 = 1,\ v_1 = 1,\ v_2 = 1$, the nonlinear algorithm changes to the linear algorithm.

the above imply that $[x^*; \mathbf{0}_{nm}]$ satisfying $(\mathbf{1}_n^\top \otimes I_m)\nabla F(x^*) = \mathbf{0}_m$ is invariant (and stable) equilibrium state of (11)-(12) for continuous-time dynamics; thus, any randomly initialized solution of $x_i$ with $y_i(0) = 0$ converges to the optimizer $\overline{x}^*$.

Table 1 summarizes the effect of the parameters $u_1,\ u_2,\ v_1,\ v_2$ on the convergence rate of the proposed signum-based dynamics (11)-(12).

**Algorithm 1.** GT-based distributed ML Algorithm

**Data:** Undirected graph topology $\mathcal{G}$, Adjacency matrices $A, W$, Loss function $f_i$
**Result:** Optimal state $x^*$
**Initialization:** $t = 0,\ y_i(0) = 0$ at all nodes and states $x_i(0)$ randomly initialized;
**While** *termination criteria NOT hold;*
**do**
Node $i$ receives $x_j$ and $y_j$ from neighbor nodes $j \in N_i$ over $\mathcal{G}$;
Node $i$ calculates $\nabla f_i(x_i)$ (the gradient of local loss function $f_i(x_i)$);
Node $i$ updates $x_i$ and $y_i$ via dynamics (11)-(12);
Node $i$ shares updated $x_i$ and $y_i$ with its neighbor nodes $i \in N_j$ over $\mathcal{G}$;

**Table 1.** The change in the parameters of signum-based nonlinearity to reach faster convergence.

| Parmeter | Faster Convergence |
|---|---|
| $0 < v_1 < 1$ | smaller $\rightarrow 0$ |
| $v_2 > 1$ | larger $\rightarrow \infty$ |
| $0 < u_1 < 1$ | smaller $\rightarrow 0$ |
| $u_2 > 1$ | larger $\rightarrow \infty$ |

The discretized version of the continuous-time dynamics (11)-(12) can be represented as follows:

$$x_i(k+1) = x_i(k) - \eta \sum_{j=1}^{n} w_{ij} \left( sgn^{v_1}\left(x_i(k) - x_j(k)\right) + sgn^{v_2}\left(x_i(k) - x_j(k)\right)\right) - \alpha y_i(k), \quad (18)$$

$$\begin{aligned} y_i(k+1) = y_i(k) - \eta \sum_{j=1}^{n} a_{ij} \left( sgn^{v_1}\left(y_i(k) - y_j(k)\right) + sgn^{v_2}\left(y_i(k) - y_j(k)\right)\right) \\ + \eta sgn^{u_1}\left(\nabla f_i(x_i(k+1)\right) - \nabla f_i(x_i(k))) + \eta sgn^{u_2}\left(\nabla f_i(x_i(k+1)\right) - \nabla f_i(x_i(k))), \end{aligned} \quad (19)$$

with $\eta$ as the discretization step-rate and $k > 0$ as the discrete step-time.

It is worth mentioning that applying the discretized version (18)-(19) may result in certain optimality gap because of the chattering phenomena. This refers to rapid and erratic oscillations in the system variables during the optimization process since the dynamics lacks Lipschitz continuity. This means that the gradients of the objective function can vary widely across different regions.

The discretization step-size plays a key role in worsening or mitigating chattering phenomena. The step-size determines how large or small the state updates we have at each iteration of the optimization algorithm. If the step-size is too large, the optimization algorithm may overshoot around the optimal solution, causing larger oscillations (and optimality gap). On the other hand, if the step-size is too small, the algorithm may converge very slowly. Thus, there is a trade-off in terms of convergence rate and steady-state optimization residual.

Table 2 summarizes the effect of the parameters $u_1,\ u_2,\ v_1,\ v_2$ on the optimality gap of the discretized dynamics (18)-(19).

**Table 2.** The change in parameters of signum-based nonlinearity to reach lower optimality gap.

| Parmeter | Faster Convergence |
|---|---|
| $0 < v_1 < 1$ | larger $\rightarrow 1$ |
| $v_2 > 1$ | smaller $\rightarrow 1$ |
| $0 < u_1 < 1$ | larger $\rightarrow 1$ |
| $u_2 > 1$ | smaller $\rightarrow 1$ |

To reduce the optimality gap introduced by non-Lipschitz signum-based update rules (18)-(19), we propose replacing a fixed step size with a diminishing step-size sequence. Diminishing step sizes (e.g., $\eta_k = \frac{\eta_0}{k+1}$ or $\eta_k = \frac{\eta_0}{\sqrt{k+1}}$) balance two competing needs: early iterations require sufficiently large updates to exploit fast transient convergence by the signum dynamics, while later iterations require progressively smaller updates to attenuate persistent bias and oscillations caused by the non-Lipschitz terms. Formally, a diminishing sequence that is positive, nonincreasing, and satisfies $\sum_{k=0}^{\infty} \eta_k = \infty$, $\sum_{k=0}^{\infty} \eta_k^2 < \infty$ preserves asymptotic convergence. In our context, this yields to vanishing optimality gap as $k \rightarrow \infty$, while on the other hand it slows the convergence. Therefore, diminishing step sizes also provide a trade-off finite-time convergence rate for improved asymptotic optimality in signum-based distributed algorithms. This is better demonstrated later by the simulations in Section 5.1.

### *4.2. Practical Implementations and Applications*

Discretizing continuous-time signum-based control laws for distributed optimization introduces several implementation challenges, mainly on stability and communication constraints. When the ideal continuous dynamics (11)-(12) is approximated with a discrete-time update, the non-Lipschitz nature of signum-based terms causes sensitivity to sampling and numerical quantization. Step-size selection in the discretized dynamics (18)-(19) trades off convergence rate against the achievable optimality gap, i.e., larger step-sizes can accelerate transient progress but increases discretization error and steady-state bias introduced by the non-Lipschitz terms and any smoothing used to avoid chattering. In particular, for signum-like updates the discretization error does not necessarily vanish with time unless the step-size is reduced; this implies that a fixed step-size gives a persistent optimality gap proportional to the step magnitude. On the other hand, diminishing step-sizes may reduce the gap asymptotically but give slow convergence and adds more complexity to the coordination of distributed agents.

The proposed signum-based distributed optimization algorithm may offer practical advantages in decentralized control and multi-agent systems where fast and robust convergence among many agents is of interest. In large-scale distributed control systems signum-like coupling may provide finite-time or very-fast convergence that help the network quickly reach consensus or track a reference despite the disturbances [60]. The non-Lipschitz nature allows stronger corrective property near disagreement regions to reduce transient errors and also improves tolerance against impulsive noise/faults. However, the trade-off between convergence speed and steady-state residual must be managed: very aggressive signum terms as in Table 1 can drive the system rapidly but introduce a non-vanishing steady-state error or chattering.

In financial data analysis and distributed machine learning on market data, signum-like functions can be used to design decentralized and robust aggregation rules, e.g., in federated learning updates resilient to outliers and deep learning models resilient to low signal-to-noise ratios [61]. For instance, signum-based penalties or gradient modifications give higher weight to correcting large discrepancies among local models or estimates as in financial chaotic systems [62]. In fact, the same non-Lipschitz behavior that accelerates convergence may prevent reaching the absolute optimal parameter set or introduce oscillations around it; therefore, hybrid designs (tempered signum terms or decaying/diminishing gains) are proposed for financial data analysis.

## 5. Simulations

*5.1. Academic Example*

For MATLAB simulation over a Core i5 Laptop, we consider a (randomly generated) dataset of $N = 100$ data points and a network of $n = 10$ agents each having access to 50% of the (randomly chosen) data points. The dataset is shown in Fig. 1. Each agent performs local regression analysis on its own batch of data and shares the regressor parameters over an Erdos-Renyi (ER) random network. The linking probability of the connected ER network is 30%. The objective function to be optimized is in the form (6)-(7). We compare the convergence under the under the nonlinear dynamics with different signum-based models (18)-(19) (as the discrete version of (11)-(12)) with $\eta = 2 \times 10^{-6}$ and $\alpha = 4$. Following Algorithm 1, agents update their regressor parameters based on the proposed dynamics and share their states $x_i$ and $y_i$ over the network. We consider four scenarios for comparison of the convergence rate and the optimality gap.

- Case (i): $u_1 = 1,\ u_2 = 1, 0 < v_1 = 0.5 < 1,\ v_2 = 1$
- Case (ii): $u_1 = 1,\ u_2 = 1, 0 < v_1 = 0.5 < 1,\ v_2 = 1.5 > 1$
- Case (iii): $0 < u_1 = 0.6 < 1,\ u_2 = 1, 0 < v_1 = 0.5 < 1,\ v_2 = 1.5 > 1$
- Case (iv): $0 < u_1 = 0.6 < 1,\ u_2 = 1.4 > 1, 0 < v_1 = 0.5 < 1,\ v_2 = 1.5 > 1$

The time evolution of the cost functions is compared in Fig. 2. As it can be seen from the figure, although as claimed in the literature the signum-based dynamics may result in finite-time stability, it results in steady-state optimization residual (depending on the parameters $u_1, u_2, v_1, v_2$).

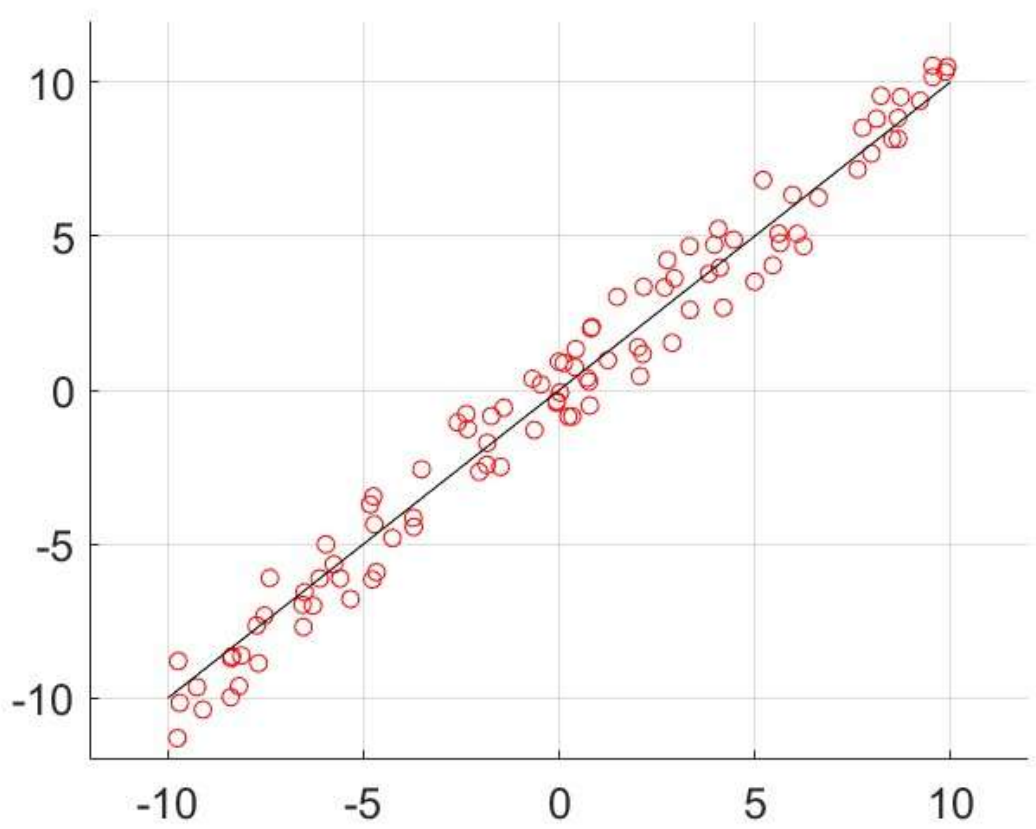

**Figure 1.** This figure shows the randomly-generated dataset and its associated regressor line to be calculated by agents in a distributed way.

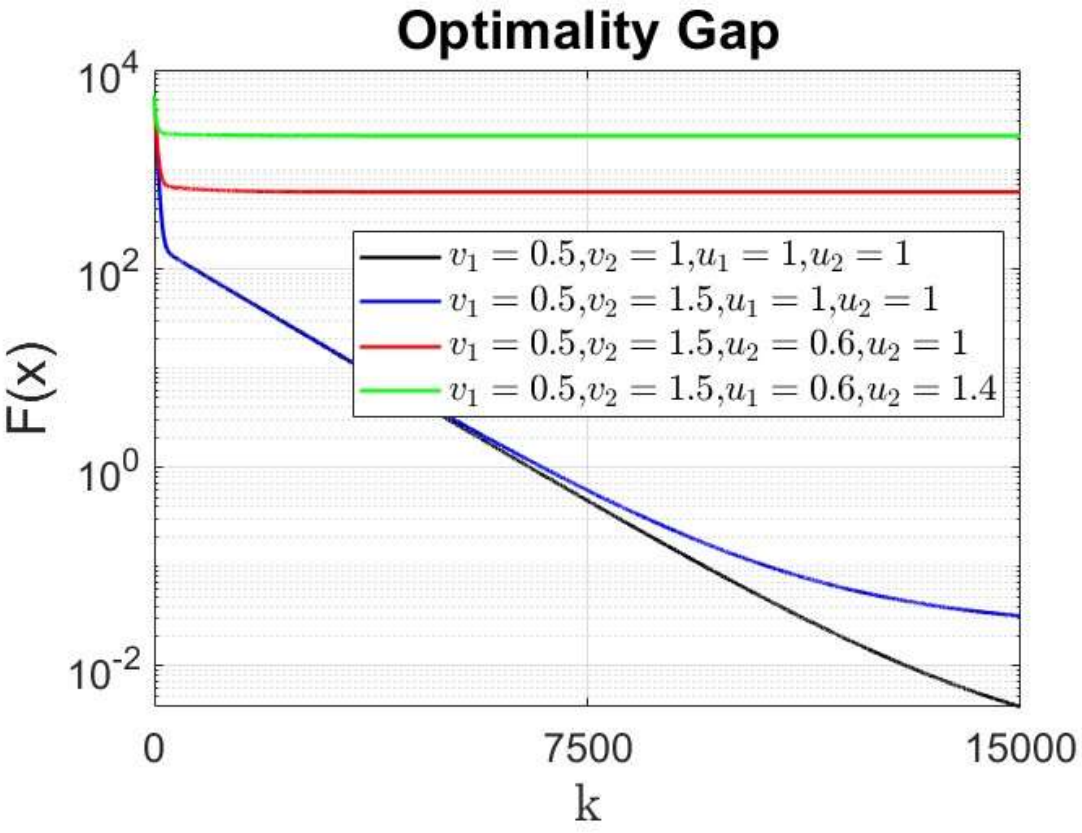

**Figure 2.** This figure compares the time evolution of DLR objective residual under different signum-based dynamics. Evidently, the discrete-time dynamics under signum-based function leads to steady-state residual.

The parameters of the regressor at different agents under different dynamics are shown in Fig. 3, Fig. 4, Fig. 5, and Fig. 6. The regressor line parameters $\beta_i, \nu_i$ are calculated at every node/agent $i$. In these figures, different colors show the parameters associated with different computing nodes/agents. As it can be seen, applying sign function may result inexact convergence and steady-state error, especially when adding it to the gradient-tracking part of the dynamics (as shown in Fig. 5 and 6).

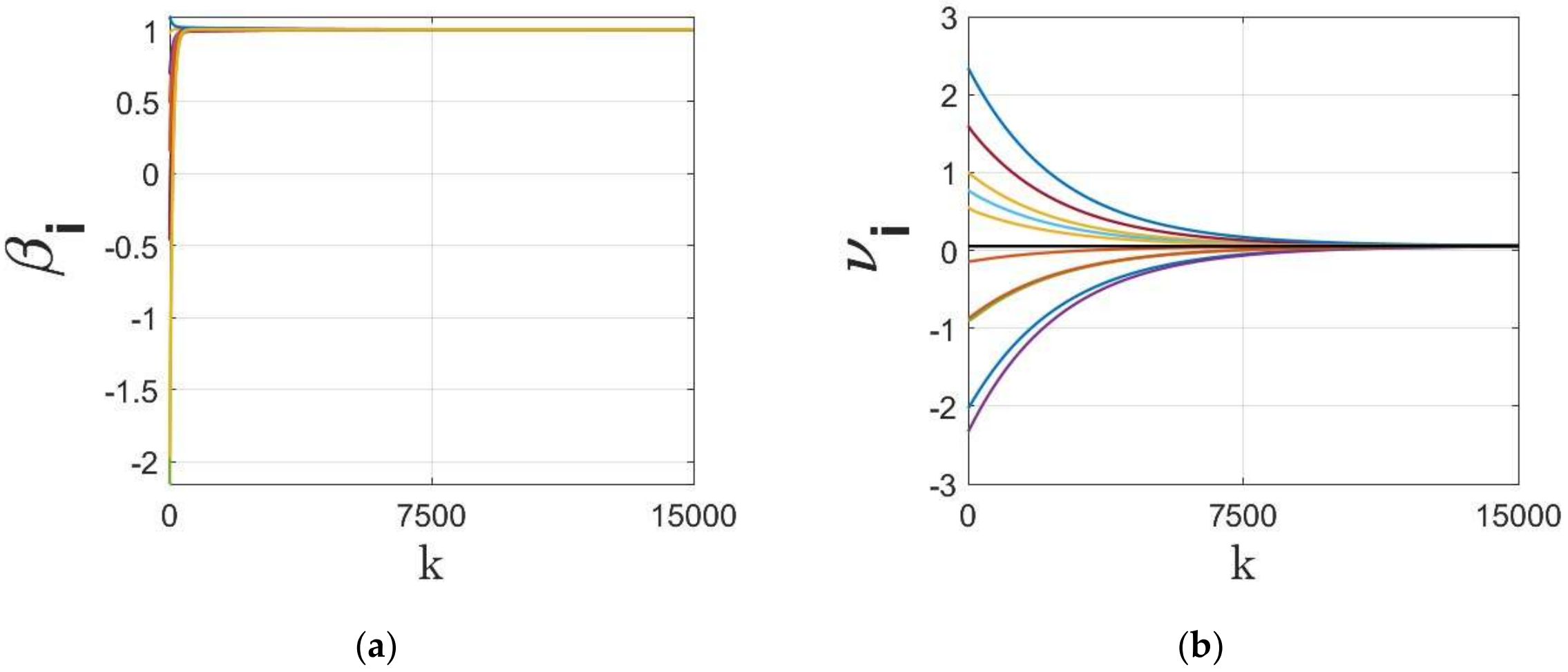

(**a**) (**b**)

**Figure 3.** This figure shows the time evolution of the regressor parameters under Case (i).

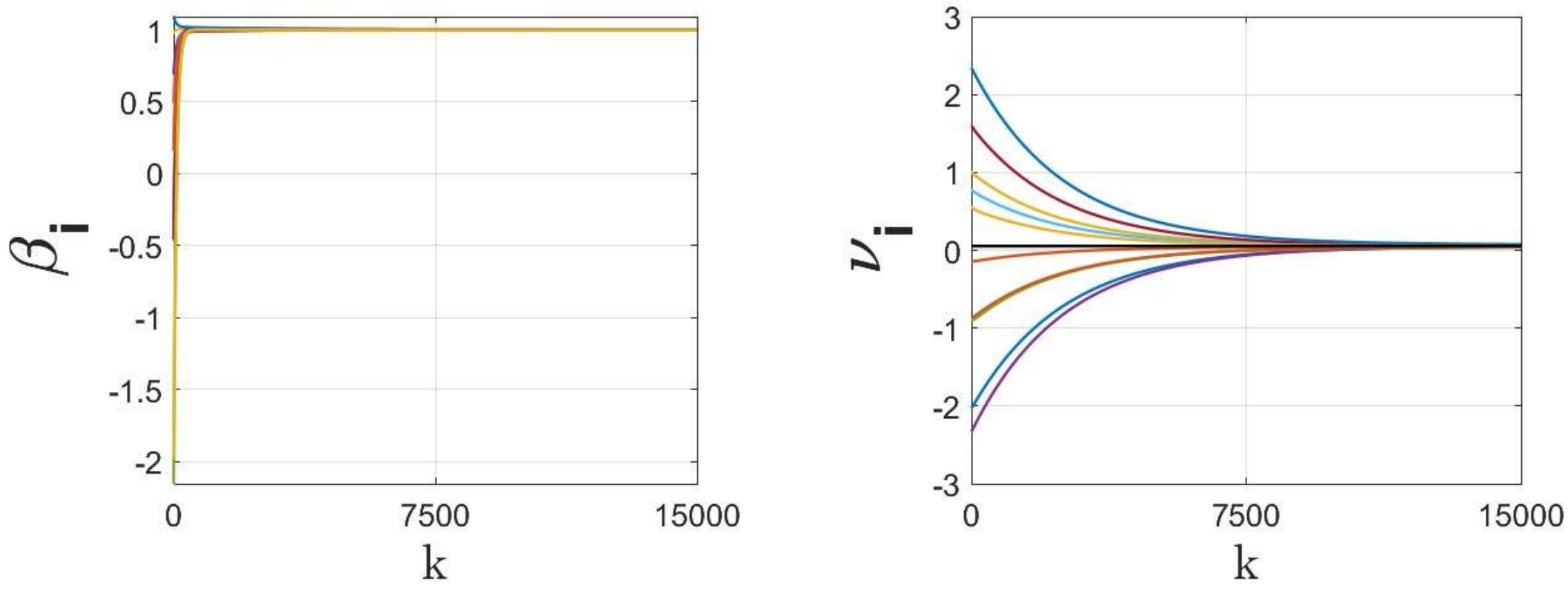

(**a**) (**b**)

**Figure 4.** This figure shows the time-evolution of the regressor parameters under Case (ii).

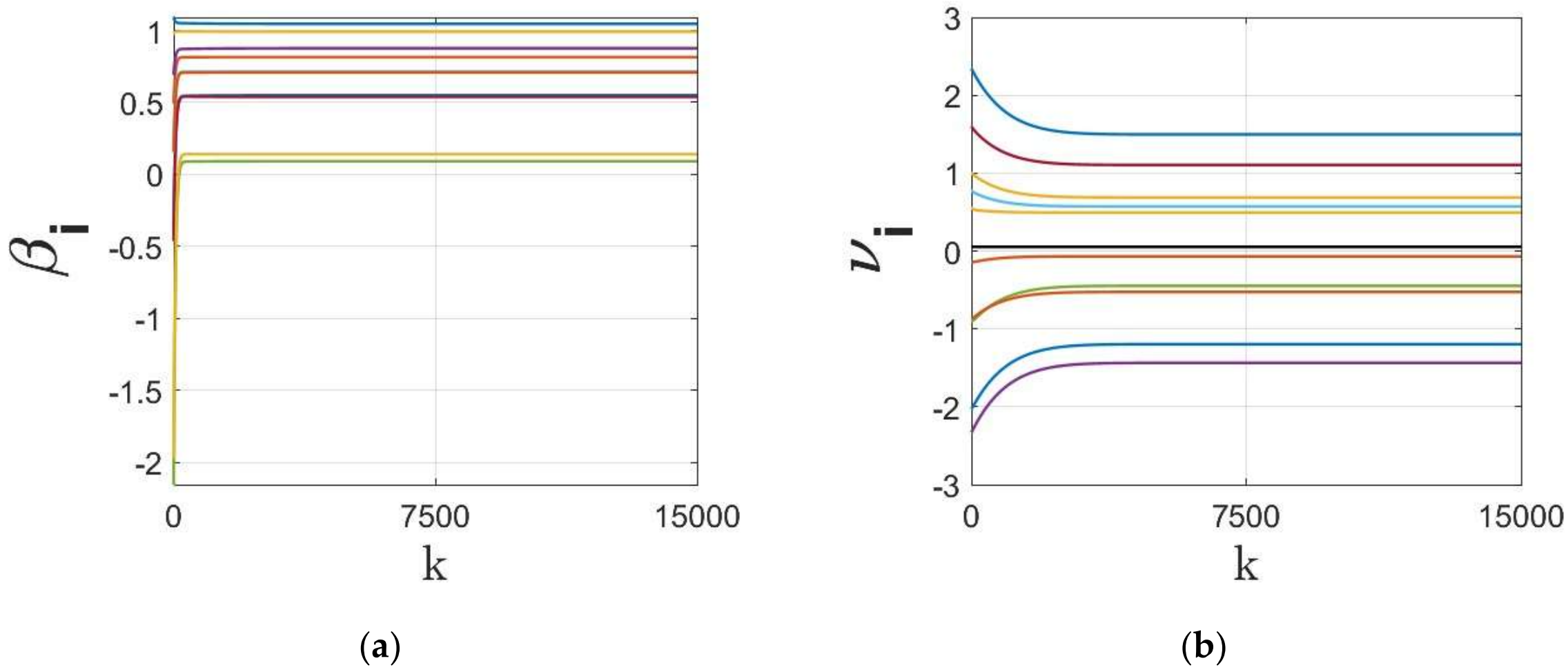

(**a**) (**b**)

**Figure 5.** This figure shows the time-evolution of the regressor parameters under Case (iii).

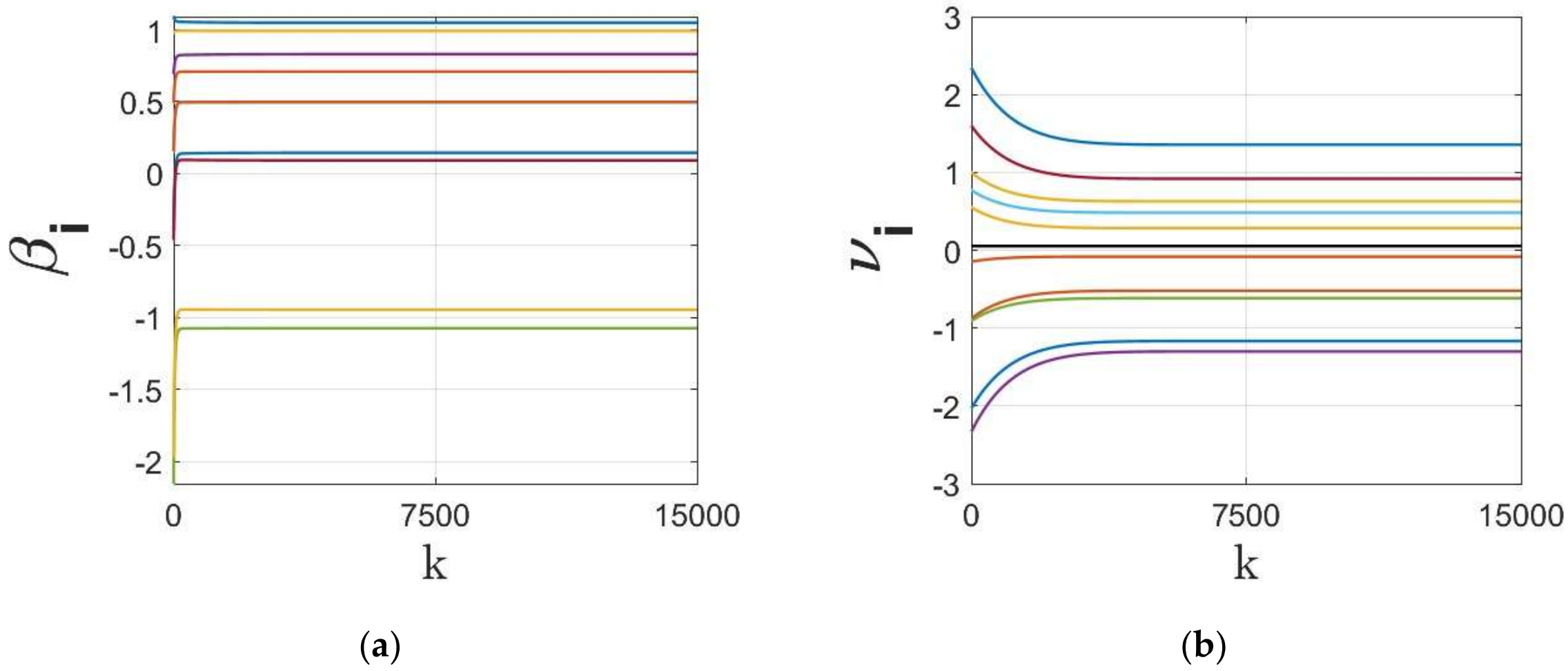

(**a**) (**b**)

**Figure 6.** This figure shows the time-evolution of the regressor parameters under Case (iv).

Next, to strengthen the generalizability of the results, we redo the simulations for large-scale example with $N = 1000$ data points and a network of $n = 100$ agents each having access to 35% of the (randomly chosen) data points. The large dataset is shown in Fig. 7. We redo the simulation on local regression with the objective function (6)-(7) over an ER random network with 20% linking probability. For this simulation, we set different values for step-size as $\eta = 1 \times 10^{-5}$ and gradient tracking rate as $\alpha = 2$. Note that, large step sizes, although may lead to faster convergence, may result in larger optimality gap of the signum-based dynamics. In case of very large step size the solution may diverge. We compare the optimality gap under four different signum-based models based on (18)-(19) as,

- Case (1): $u_1 = 1,\ u_2 = 1, 0 < v_1 = 0.3 < 1,\ v_2 = 1$
- Case (2): $u_1 = 1,\ u_2 = 1, 0 < v_1 = 0.3 < 1,\ v_2 = 2 > 1$
- Case (3): $0 < u_1 = 0.8 < 1,\ u_2 = 1, 0 < v_1 = 0.3 < 1,\ v_2 = 2 > 1$
- Case (4): $0 < u_1 = 0.8 < 1,\ u_2 = 1.2 > 1, 0 < v_1 = 0.3 < 1,\ v_2 = 2 > 1$

The residual cost function over iteration $k$ is compared in Fig. 8. Despite the finite-time convergence, the solution may result in steady-state residual or optimality gap depending on the parameters $u_1$, $u_2$, $v_1$, $v_2$. To better highlight this optimality gap on the regressor line parameters$\beta_i, \nu_i$, these parameters are shown in Fig. 9, Fig. 10, Fig. 11, and Fig. 12 for different cases under signum-based dynamics. It is clear that applying sign function may result in steady-state error, especially for $u_1 \neq 1$, $u_2 \neq 1$ where the gradient-tracking is under signum-based nonlinearity (e.g., see Fig. 11 and 12).

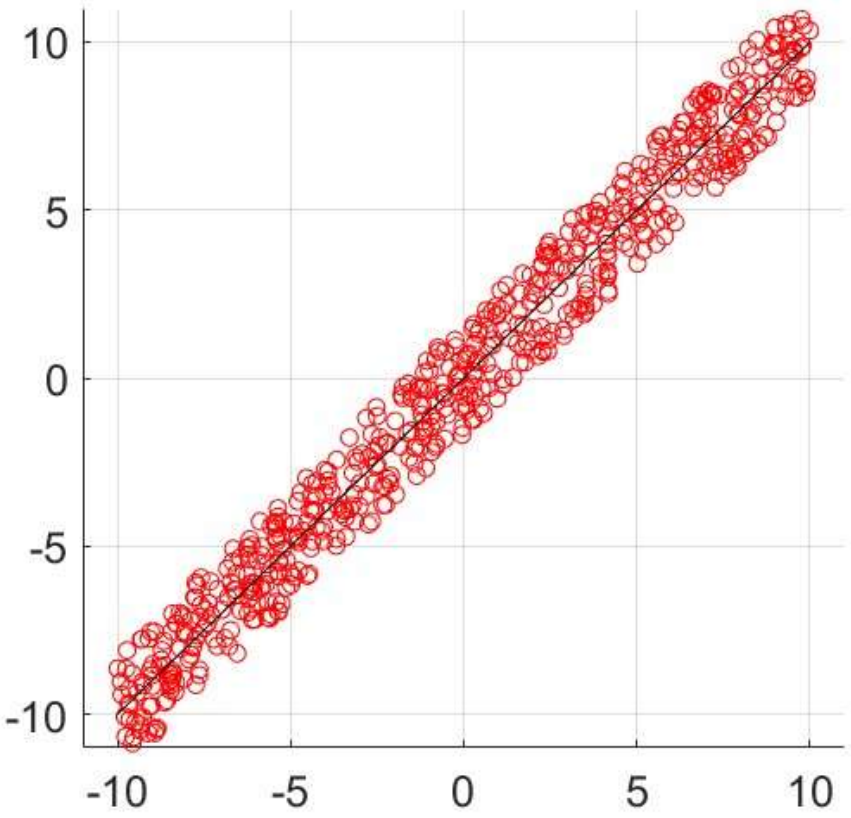

**Figure 7.** This figure shows the randomly-generated data points with the associated regressor line for the large-scale distributed optimization simulation.

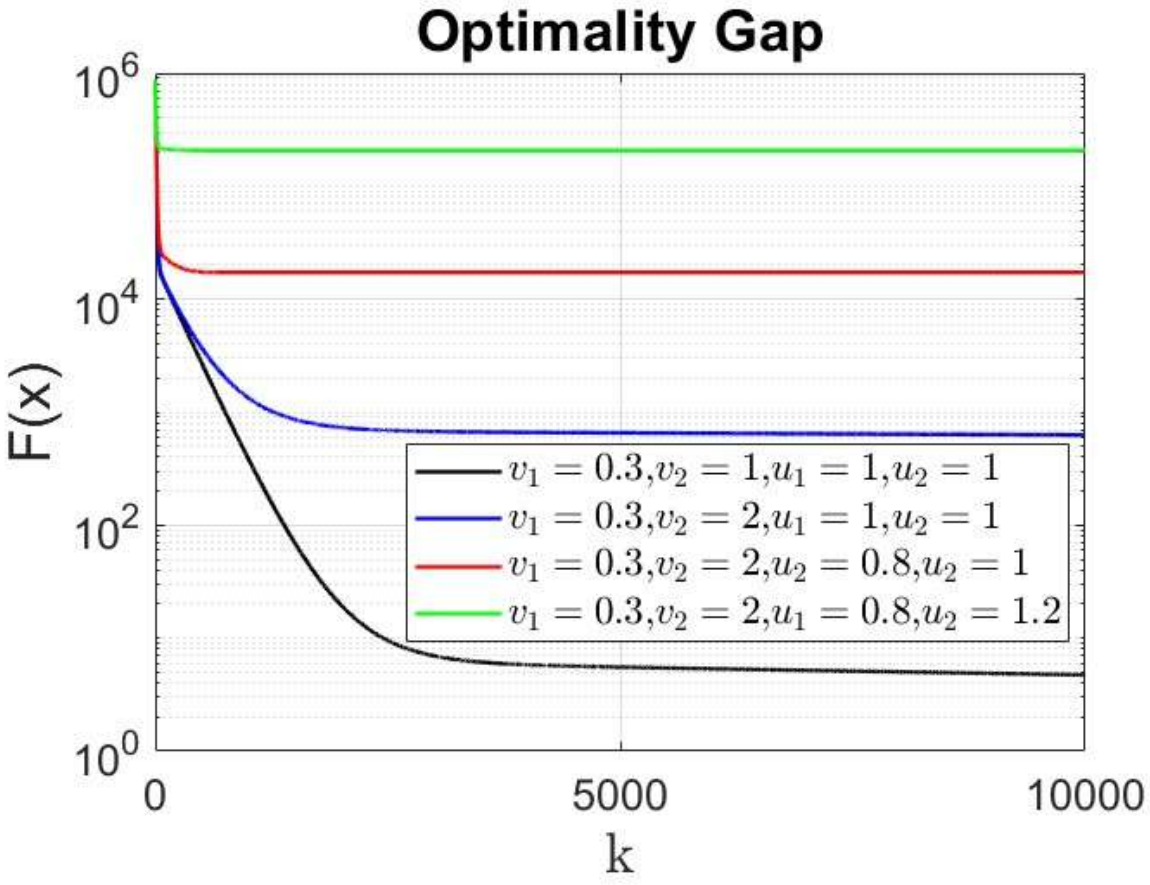

**Figure 8.** This figure compares the DLR optimality gap under different signum-based dynamics where signum-based solution may result in some optimality gap.

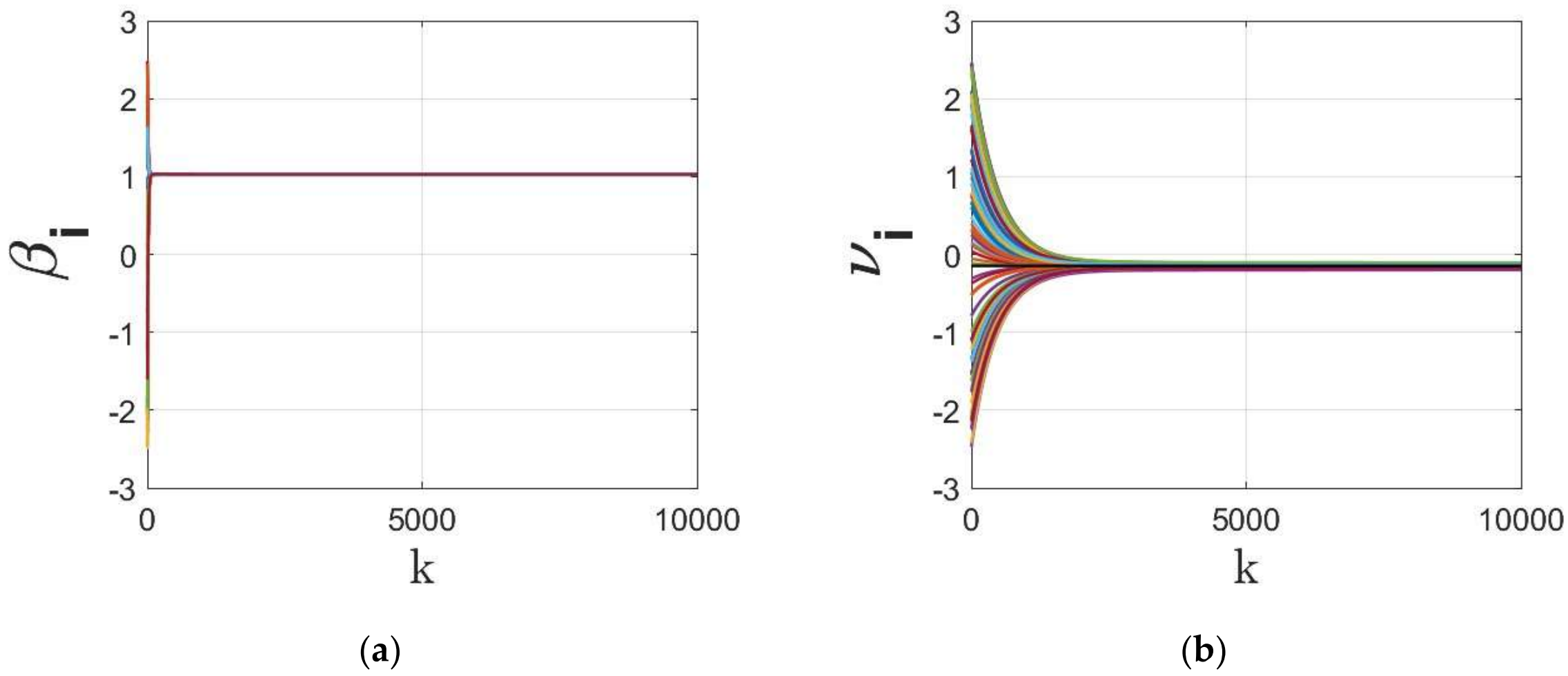

**Figure 9.** This figure shows the time evolution of the regressor parameters under Case (1).

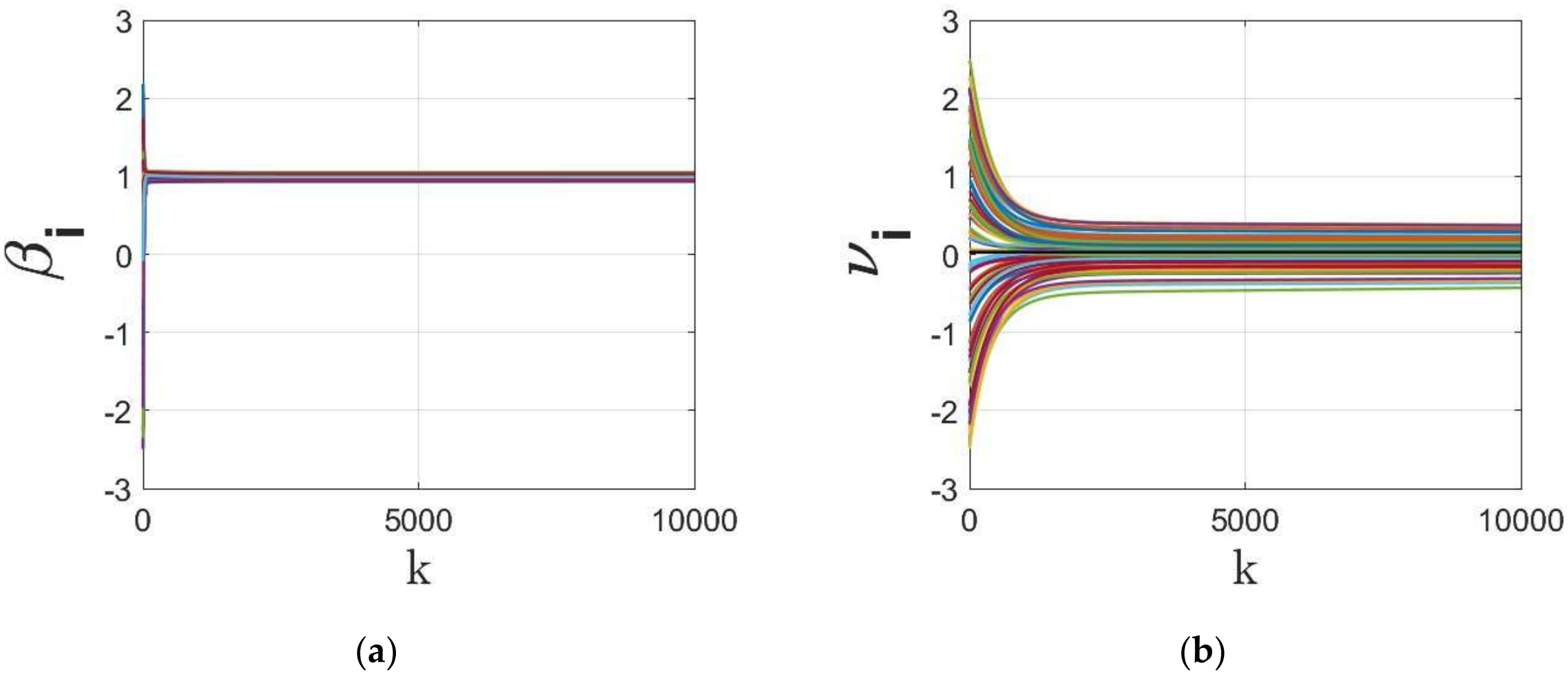

**Figure 10.** This figure shows the time evolution of the regressor parameters under Case (2).

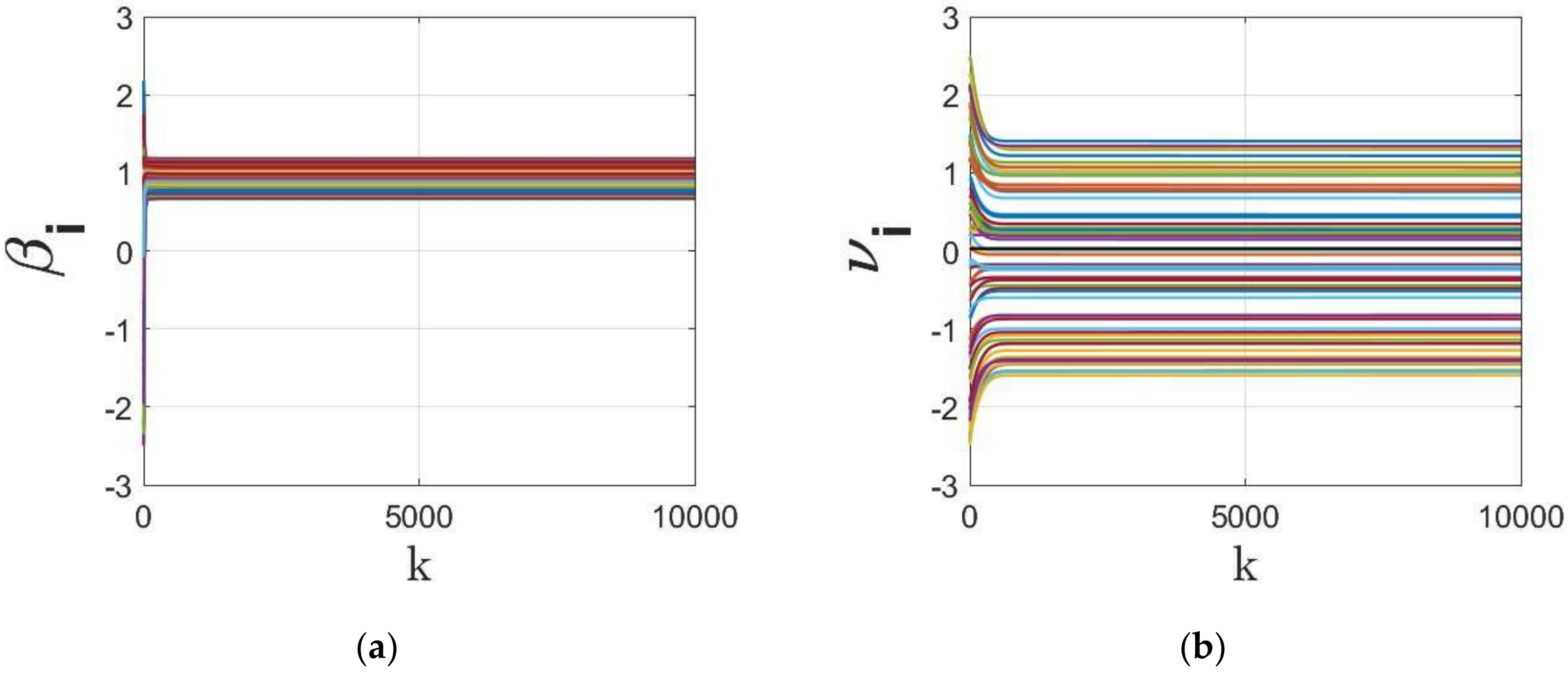

**Figure 11.** This figure shows the time evolution of the regressor parameters under Case (3).

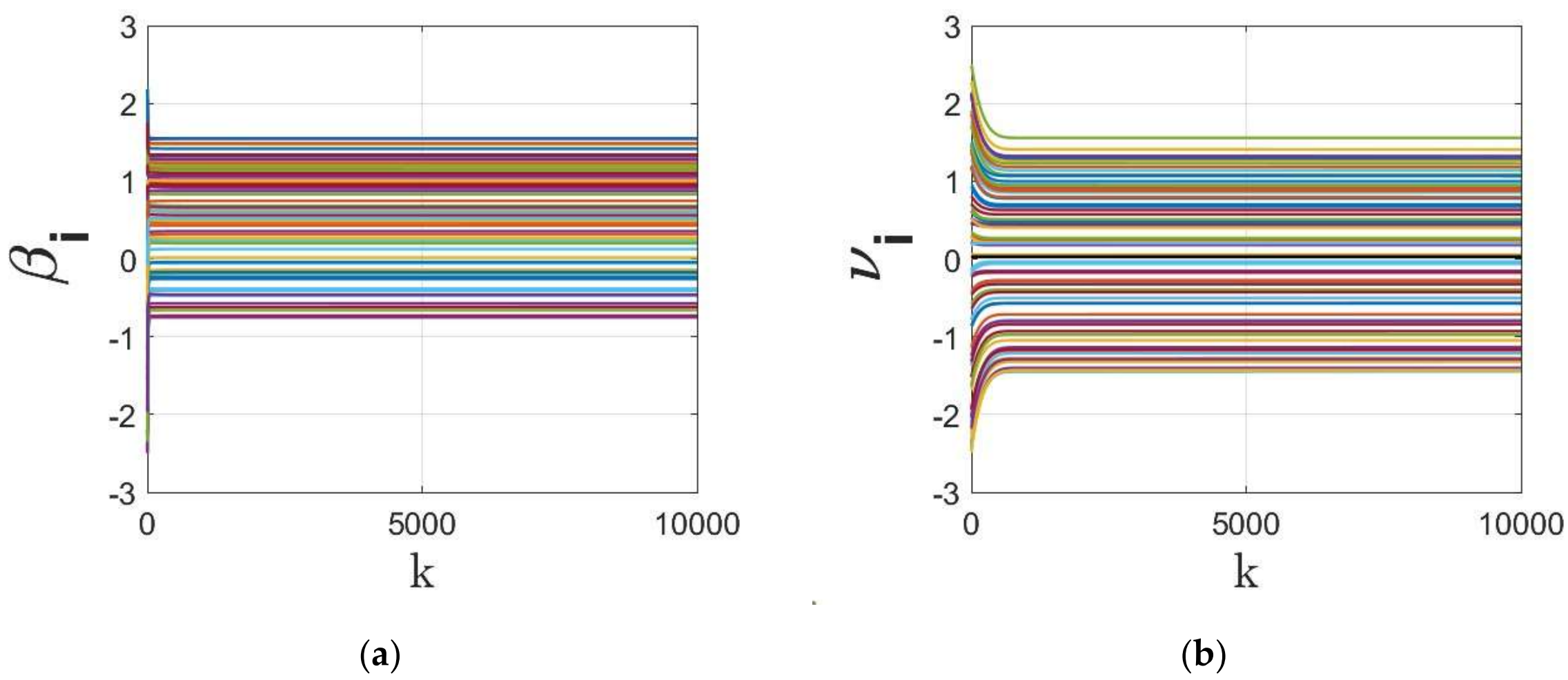

(**a**) (**b**)

**Figure 12.** This figure shows the time evolution of the regressor parameters under Case (4).

Next, we repeat the simulation to compare the optimality gap under fixed and diminishing step sizes η. For fixed step size we consider $\eta = 5 \times 10^{-6}$ and for diminishing step size $\eta = \frac{5\times 10^{-5}}{\sqrt{k+1}}$. The signum function parameters are set as $u_1 = 1,\ u_2 = 1,\ 0 < v_1 = 0.75 < 1,\ v_2 = 1.25 > 1$. As we see from Fig. 13, the optimality gap is smaller for diminishing step size as compared with the fixed step size, while converging in slower rate. This is one remedy to decrease the optimality gap while it causes slower convergence rate.

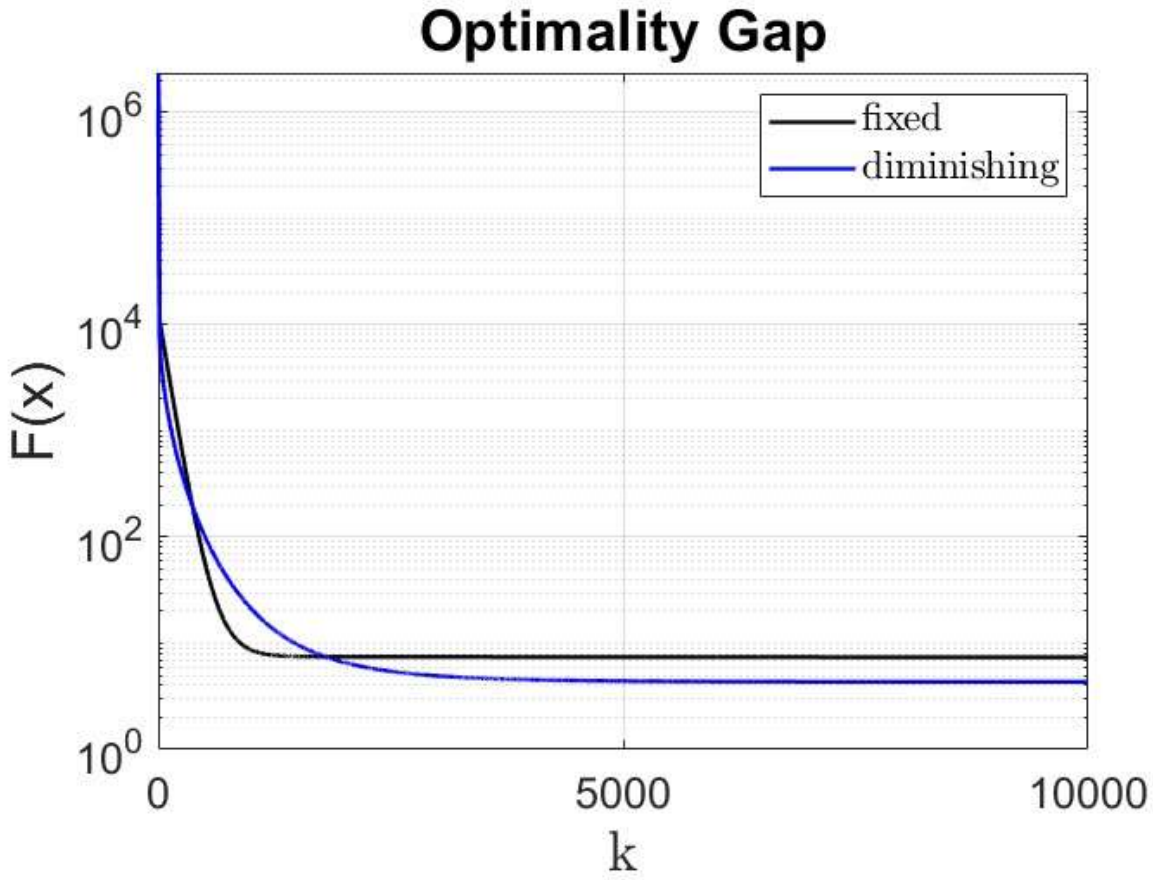

**Figure 13.** This figure compares the DLR optimality gap under fixed step size and diminishing step size.

*5.2. Real Dataset Example*

For the next simulation, we consider the MNIST dataset and the optimization data from [63]. We randomly select $N = 12000$ labelled images from this dataset and classify these images via logistic regression with a convex regularizer over an exponential network of $n = 16$ computing nodes. For this problem, the global cost function is

$$\min_{b,c} F(b,c) = \frac{1}{n}\sum_{i=1}^{n} f_i\,, \quad (20)$$

with each computing node $i$ taking a batch of $m_i = 750$ sample images. Then, every node $i$ locally minimizes the following objective function:

$$f_i(b,c) = \frac{1}{m_i}\sum_{j=1}^{m_i} \ln\left(1 + exp\left(-\left(b^\top x_{i,j} + c\right)y_{i,j}\right)\right) + \frac{\lambda}{2}\left||b|\right|_2^2, \tag{21}$$

where $b, c$ denote the parameters of the separating hyperplane for classification. The optimization residual (or the optimality gap) of the signum-based Algorithm 1 with $u_1 = 1, u_2 = 1, 0 < v_1 = 0.9 < 1, v_2 = 1.1 > 1$ is compared with some existing algorithms in the literature. The following algorithms are considered for comparison: GP [64], SGP [65], S-ADDOPT [66]. The comparison results are shown in Fig. 14. As it is clear from the figure, by choosing $v_1$, and $v_2$ moderately close to 1, the Algorithm 1 reaches fast convergence with sufficiently low optimality-gap.

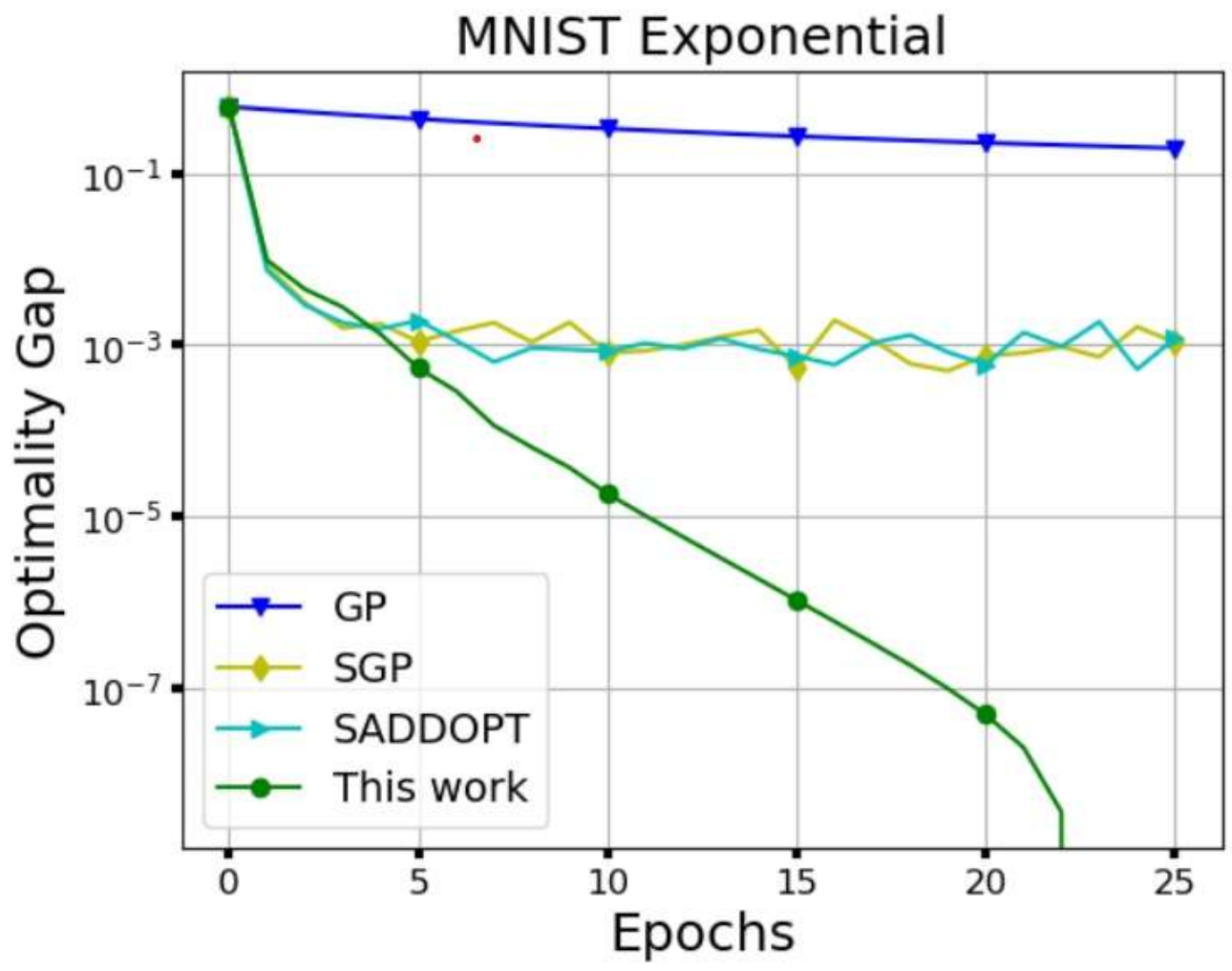

**Figure 14.** This figure compares the performance of the signum-based algorithm with some existing algorithms in the literature.

## 6. Conclusions

### *6.1. Concluding Remarks*

Our findings in this paper suggest that, although non-Lipschitz signum-based functions show interesting behavior in terms of convergence time/rate, they suffer from certain optimality gap in discrete-time applications. The use of such functions results in losing the exact optimality while, on the other hand, reaching rapid convergence rates. In other words, such solutions are practical in scenarios where the exact convergence can be given up to reach faster convergence. This trade-off presents a valuable consideration to balance between the need for quick convergence with the required optimality of the solutions in distributed optimization and machine learning applications.

### *6.2. Future Directions*

The results can be extended to other applications, to study the trade-off between convergence rate and optimality gap in distributed optimization and learning algorithms via alternating direction method of multipliers (D-ADMM) [67] and in the distributed control of the integrated energy system [68]. As we move forward, studying the integration of non-Lipschitz signum-based functions into distributed/decentralized frameworks is an interesting method for other multi-agent applications. This study provides practical insights for decision-making algorithms in real-world applications, for example, distributed techniques for estimation, detection, and resource allocation.

**Author Contributions:** Conceptualization, A.A., Z.R.G, H.R.R. and M.D.; methodology, A.A., Z.R.G, H.R.R. and M.D.; software, M.D.; validation, M.D.; formal analysis, M.D.; investigation, A.A., Z.R.G, H.R.R. and M.D.; resources, M.D.; data curation, M.D.; writing—original draft preparation, A.A.G., Z.R.G., and M.D.; writing—review and editing, A.A.G., A.A., Z.R.G, H.R.R. and M.D.; visualization, M.D.; supervision, M.D.; project administration, M.D. All authors have read and agreed to the published version of the manuscript.

**Funding:** This work has been supported by the Center for International Scientific Studies & Collaborations (CISSC), Ministry of Science, Research and Technology of Iran.

**Data Availability Statement:** The MNIST data set is available at the following link: https://www.kaggle.com/datasets/hojjatk/mnist-dataset. The raw data on MATLAB simulations supporting the conclusions of this article will be made available by the authors on request.

**Conflicts of Interest:** The authors declare no conflicts of interest.

## References

1. I. Necoara, V. Nedelcu, and I. Dumitrache, "Parallel and distributed optimization methods for estimation and control in networks," Journal of Process Control, vol. 21, no. 5, pp. 756–766, 2011.
2. A. Verma and S. Butenko, "A distributed approximation algorithm for the bottleneck connected dominating set problem," Optimization Letters, vol. 6, no. 8, pp. 1583–1595, 2012.
3. A. Veremyev, V. Boginski, and E. L. Pasiliao, "Potential energy principles in networked systems and their connections to optimization problems on graphs," Optimization Letters, vol. 9, pp. 585–600, 2015.
4. M. I. Qureshi, A. I. Rikos, T. Charalambous, U. A. Khan, "Learning and Optimization in Wireless Sensor Networks," Wireless Sensor Networks in Smart Environments: Enabling Digitalization from Fundamentals to Advanced Solutions, vol. 10, pp. 35-64, 2025.
5. Z. R. Gabidullina, "Design of the best linear classifier for box-constrained data sets," in Mesh Methods for Boundary-Value Problems and Applications: 13th International Conference, Kazan, Russia. Springer, 2021, pp. 109–124.
6. J. Alzubi, A. Nayyar, and A. Kumar, "Machine learning from theory to algorithms: an overview," in Journal of physics: conference series. IOP Publishing, 2018, vol. 1142, p. 012012.
7. M. Doostmohammadian, A. Aghasi, T. Charalambous, and U. A. Khan, "Distributed support vector machines over dynamic balanced directed networks," IEEE Control Systems Letters, vol. 6, pp. 758 – 763, 2021.
8. L. Cui, S. Yang, F. Chen, Z. Ming, N. Lu, and J. Qin, "A survey on application of machine learning for internet of things," International Journal of Machine Learning and Cybernetics, vol. 9, pp. 1399–1417, 2018.
9. H. Fourati, R. Maaloul, and L. Chaari, "A survey of 5g network systems: challenges and machine learning approaches," International Journal of Machine Learning and Cybernetics, vol. 12, pp. 385–431, 2021.
10. Z. Xie and Z. Wu, "Event-triggered consensus control for dc microgrids based on MKELM and state observer against false data injection attacks," International Journal of Machine Learning and Cybernetics, vol. 15, pp. 775–793, 2024.
11. M. Doostmohammadian, H. R. Rabiee, and U. A. Khan, "Cyber-social systems: modeling, inference, and optimal design," IEEE Systems Journal, vol. 14, no. 1, pp. 73–83, 2019.
12. S. S. Stanković, M. Beko, and M. S. Stanković, "Nonlinear robustified stochastic consensus seeking," Systems & Control Letters, vol. 139, pp. 104667, 2020.
13. D. Jakovetic, M. Vukovic, D. Bajovic, A. K. Sahu, and S. Kar, "Distributed recursive estimation under heavy-tail communication noise," SIAM Journal on Control and Optimization, vol. 61, no. 3, pp. 1582–1609, 2023.
14. L. Dai, X. Chen, L. Guo, J. Zhang, and J. Chen, "Prescribed-time group consensus for multiagent system based on a distributed observer approach," International Journal of Control, Automation and Systems, vol. 20, no. 10, pp. 3129–3137, 2022.
15. B. Zhang, S. Mo, H. Zhou, T. Qin, and Y. Zhong, "Finite-time consensus tracking control for speed sensorless multi-motor systems," Applied Sciences, vol. 12, no. 11, pp. 5518, 2022.
16. M. Doostmohammadian, "Single-bit consensus with finite-time convergence: Theory and applications," IEEE Transactions on Aerospace and Electronic Systems, vol. 56, no. 4, pp. 3332–3338, 2020.
17. C. Ge, L. Ma, and S. Xu, "Distributed fixed-time leader-following consensus for multi-agent systems: an event-triggered mechanism," actuators, vol. 13, no. 1, pp. 40, 2024.

18. J. Yang, R. Li, Q. Gan, and X. Huang, "Zero-sum-game-based fixed-time event-triggered optimal consensus control of multi-agent systems under FDI attacks," Mathematics, vol. 13, no. 3, pp. 543, 2025.
19. M. Doostmohammadian, A. Aghasi, M. Pirani, E. Nekouei, U. A. Khan, and T. Charalambous, "Fast-convergent anytime-feasible dynamics for distributed allocation of resources over switching sparse networks with quantized communication links," in IEEE European Control Conference, 2022, pp. 84–89.
20. M. Doostmohammadian and N. Meskin, "Finite-time stability under denial of service," IEEE Systems Journal, vol. 15, no. 1, pp. 1048–1055, 2020.
21. P. Budhraja, M. Baranwal, K. Garg, and A. Hota, "Breaking the convergence barrier: Optimization via fixed-time convergent flows," in Proceedings of the AAAI Conference on Artificial Intelligence, 2022, vol. 36, pp. 6115–6122.
22. J. Mishra and X. Yu, "On fixed-time convergent sliding mode control design and applications," Emerging Trends in Sliding Mode Control: Theory and Application, pp. 203–237, 2021.
23. H. Ríos, D. Efimov, J. A Moreno, W. Perruquetti, and J. G. Rueda- Escobedo, "Time-varying parameter identification algorithms: Finite and fixed-time convergence," IEEE Transactions on Automatic Control, vol. 62, no. 7, pp. 3671–3678, 2017.
24. G. Malaspina, D. Jakovetić, and N. Krejić, "Linear convergence rate analysis of a class of exact first-order distributed methods for weight-balanced time-varying networks and uncoordinated step sizes," Optimization Letters, pp. 1–22, 2023.
25. Z. Zuo, Q. Han, and B. Ning, Fixed-time cooperative control of multi-agent systems, Springer, 2019.
26. L. Acho, P. Buenestado, and G. Pujol, "A finite-time control design for the discrete-time chaotic logistic equations," Actuators, vol. 13, no. 8, pp. 295, 2024.
27. M. Basin, "Finite-and fixed-time convergent algorithms: Design and convergence time estimation," Annual reviews in control, vol. 48, pp. 209–221, 2019.
28. C. Xi and U. A. Khan, "Distributed subgradient projection algorithm over directed graphs," IEEE Transactions on Automatic Control, vol. 62, no. 8, pp. 3986–3992, 2016.
29. Z. Szabó, B. K. Sriperumbudur, B. Póczos, and A. Gretton, "Learning theory for distribution regression," The Journal of Machine Learning Research, vol. 17, no. 1, pp. 5272–5311, 2016.
30. R. Xin, U. Khan, and S. Kar, "A hybrid variance-reduced method for decentralized stochastic non-convex optimization," in International Conference on Machine Learning. PMLR, 2021, pp. 11459–11469.
31. B. Du, J. Zhou, and D. Sun, "Improving the convergence of distributed gradient descent via inexact average consensus," Journal of Optimization Theory and Applications, vol. 185, pp. 504–521, 2020.
32. A. Simonetto and H. Jamali-Rad, "Primal recovery from consensus-based dual decomposition for distributed convex optimization," Journal of Optimization Theory and Applications, vol. 168, pp. 172–197, 2016.
33. Z. Liu, H. Jahanshahi, C. Volos, S. Bekiros, S. He, M. O. Alassafi, and A. M. Ahmad, "Distributed consensus tracking control of chaotic multi-agent supply chain network: A new fault-tolerant, finite-time, and chatter-free approach," Entropy, vol. 24, no. 1, pp. 33, 2021.
34. Z. Yu, S. Yu, H. Jiang, and X. Mei, "Distributed fixed-time optimization for multi-agent systems over a directed network," Nonlinear Dynamics, vol. 103, pp. 775–789, 2021.
35. X. Shi, G. Wen, and X. Yu, "Finite-time convergent algorithms for time-varying distributed optimization," IEEE Control Systems Letters, 2023.
36. W. Tang and P. Daoutidis, "Fast and stable nonconvex constrained distributed optimization: the ellada algorithm," Optimization and Engineering, vol. 23, no. 1, pp. 259–301, 2022.
37. S. Li, X. Nian, Z. Deng, and Z. Chen, "Predefined-time distributed optimization of general linear multi-agent systems," Information Sciences, vol. 584, pp. 111–125, 2022.
38. X. Gong, Y. Cui, J. Shen, J. Xiong, and T. Huang, "Distributed optimization in prescribed-time: Theory and experiment," IEEE Transactions on Network Science and Engineering, vol. 9, no. 2, pp. 564–576, 2021.
39. C. Deng, M.g Ge, Z. Liu, and Y. Wu, "Prescribed-time stabilization and optimization of cps-based microgrids with event-triggered interactions," International Journal of Dynamics and Control, pp. 1–13, 2023.
40. X. Wen and S. Qin, "A projection-based continuous-time algorithm for distributed optimization over multi-agent systems," Complex & Intelligent Systems, vol. 8, no. 2, pp. 719–729, 2022.
41. Q. Li, M. Wang, H. Sun, and S. Qin, "An adaptive finite-time neurodynamic approach to distributed consensus-based optimization problem," Neural Computing and Applications, vol. 35, no. 28, pp. 20841–20853, 2023.
42. J.J. Slotine and W. Li, Applied nonlinear control, Prentice-Hall, 1991.

43. H. Wu, F. Tao, L. Qin, R. Shi, and L. He, “Robust exponential stability for interval neural networks with delays and non-lipschitz activation functions,” Nonlinear Dynamics, vol. 66, no. 4, pp. 479– 487, 2011.
44. H. Yu and H. Wu, “Global robust exponential stability for hopfield neural networks with non-lipschitz activation functions,” Journal of Mathematical Sciences, vol. 187, no. 4, pp. 511–523, 2012.
45. M. Balcan, A. Blum, D. Sharma, and H. Zhang, “An analysis of robustness of non-lipschitz networks,” Journal of Machine Learning Research, vol. 24, no. 98, pp. 1–43, 2023.
46. W. Li, W. Bian, and X. Xue, “Projected neural network for a class of non-lipschitz optimization problems with linear constraints,” IEEE Transactions on Neural Networks and Learning Systems, vol. 31, no. 9, pp. 3361–3373, 2019.
47. R. Olfati-Saber and R. M. Murray, “Consensus problems in networks of agents with switching topology and time-delays,” IEEE Transactions on Automatic Control, vol. 49, no. 9, pp. 1520–1533, Sept. 2004.
48. F. Wang, Y. Zhang, L. Zhang, J. Zhang, and Y. Huang, “Finite-time consensus of stochastic nonlinear multi-agent systems,” International Journal of Fuzzy Systems, vol. 22, pp. 77–88, 2020.
49. B. Zhang and Y. Jia, “Fixed-time consensus protocols for multi-agent systems with linear and nonlinear state measurements,” Nonlinear Dynamics, vol. 82, pp. 1683–1690, 2015.
50. J. Liu, Y. Yu, Q. Wang, and C. Sun, “Fixed-time event-triggered consensus control for multi-agent systems with nonlinear uncertainties,” Neurocomputing, vol. 260, pp. 497–504, 2017.
51. Z. Zuo, B. Tian, M. Defoort, and Z. Ding, “Fixed-time consensus tracking for multiagent systems with high-order integrator dynamics,” IEEE Transactions on Automatic Control, vol. 63, no. 2, pp. 563–570, 2017.
52. K. Yan, T. Han, B. Xiao, and H. Yan, “Distributed fixed-time and prescribed-time average consensus for multi-agent systems with energy constraints,” Information Sciences, vol. 647, pp. 119471, 2023.
53. Y. Yu, C. Liu, Y. Li, and H. Li, “Practical fixed-time distributed average-tracking with input delay based on event-triggered method,” International Journal of Control, Automation and Systems, vol. 21, no. 3, pp. 845–853, 2023.
54. M. Doostmohammadian, S. Kharazmi, and H. R. Rabiee, “How clustering affects the convergence of decentralized optimization over networks: a monte-carlo-based approach,” Social Network Analysis and Mining, vol. 14, no. 1, pp. 135, 2024.
55. M. Doostmohammadian and A. Aghasi, “Accelerated distributed allocation,” IEEE Signal Processing Letters, 2024.
56. S. Boyd, N. Parikh, E. Chu, B. Peleato, and J. Eckstein, "Distributed optimization and statistical learning via the alternating direction method of multipliers," Foundations and Trends® in Machine learning, vol. 3, no. 1, pp. 1-122, 2011.
57. C. Song, S. Yoon, and V. Pavlovic, “Fast ADMM algorithm for distributed optimization with adaptive penalty,” in Proceedings of the AAAI Conference on Artificial Intelligence, 2016, vol. 30.
58. Z. Lin, H. Li, and C. Fang, “ADMM for distributed optimization,” in Alternating Direction Method of Multipliers for Machine Learning, pp. 207–240. Springer, 2022.
59. M. Ma, A. N. Nikolakopoulos, and G. B. Giannakis, “Hybrid admm: a unifying and fast approach to decentralized optimization,” EURASIP Journal on Advances in Signal Processing, vol. 2018, pp. 1–17, 2018.
60. M. Gautam, A. Pati, S. Mishra, B. Appasani, E. Kabalci, N. Bizon, and P. Thounthong, “A comprehensive review of the evolution of networked control system technology and its future potentials,” Sustainability, vol. 13, no. 5, pp. 2962, 2021.
61. L. Zhang and L. Hua, “Major issues in high-frequency financial data analysis: A survey of solutions,” Mathematics, vol. 13, no. 3, pp. 347, 2025.
62. Y. Wei, C. Xie, X. Qing, and Y. Xu, “Control of a new financial risk contagion dynamic model based on finite-time disturbance,” Entropy, vol. 26, no. 12, pp. 999, 2024.
63. M. I. Qureshi and U. A. Khan, “Stochastic first-order methods over distributed data,” in IEEE 12th Sensor Array and Multichannel Signal Processing Workshop (SAM), 2022, pp. 405–409.
64. A. Nedić and A. Olshevsky, “Distributed optimization over time-varying directed graphs,” IEEE Transactions on Automatic Control, vol. 60, no. 3, pp. 601–615, 2014.
65. A. Spiridonoff, A. Olshevsky, and I. Paschalidis, “Robust asynchronous stochastic gradient-push: Asymptotically optimal and network-independent performance for strongly convex functions,” Journal of Machine Learning Research, vol. 21, no. 58, pp. 1–47, 2020.
66. M. I. Qureshi, R. Xin, S. Kar, and U. A. Khan, “S-ADDOPT: Decentralized stochastic first-order optimization over directed graphs,” IEEE Control Systems Letters, vol. 5, no. 3, pp. 953–958, 2020.
67. Y. Noah and N. Shlezinger, “Distributed learn-to-optimize: Limited communications optimization over networks via deep unfolded distributed ADMM,” IEEE Transactions on Mobile Computing, vol. 24, no. 4, pp. 3012–3024, 2025.

68. N. Zhang, Q. Sun, L. Yang, and Y. Li, "Event-triggered distributed hybrid control scheme for the integrated energy system," IEEE Transactions on Industrial Informatics, vol. 18, no. 2, pp. 835–846, 2022.